\documentclass[12pt]{article}
\usepackage{graphics}
\usepackage{bm}
\usepackage{graphicx}
\usepackage{amsmath}
\usepackage{amssymb}
\usepackage{amstext}
\usepackage{amscd}
\usepackage{amsfonts}
\usepackage{appendix}
\usepackage{color}
\DeclareMathAlphabet{\EuFrak}{U}{euf}{m}{n}
\DeclareMathAlphabet{\EuScript}{U}{eus}{m}{n}

\newcommand{\nd}{\noindent}

\newcommand{\be}{\begin{equation}}
\newcommand{\ee}{\end{equation}}
\newcommand{\ben}{\begin{eqnarray}}
\newcommand{\een}{\end{eqnarray}}

\title{{\bf Hypergeometric Connotations of Quantum
Equations}}
\author{{A. Plastino$^1$, M.C.Rocca$^1$} \\
\small{$^1$ La Plata National University and
Argentina's National Research Council}\\
\small{(IFLP-CCT-CONICET)-C. C. 727, 1900 La Plata - Argentina}}

\date{\today}

\begin{document}

\maketitle

\begin{abstract}
\nd We show that the Schr\"{o}dinger and Klein-Gordon equations
can both be derived from an Hypergeometric differential equation.
The same applies to non linear generalizations of these
equations.\vskip 3mm

\nd Keywords: Schr\"{o}dinger equation,  Klein-Gordon equation,
hypergeometric functions.

\end{abstract}

\newpage

\renewcommand{\theequation}{\arabic{section}.\arabic{equation}}

\section{Introduction}

\setcounter{equation}{0}

\nd The time-dependent Schr\"{o}dinger  equation is one of the
more basic ones  of physics, ruling all phenomena of the
microscopic world and some of the macro. This fact
notwithstanding, its origin is  not yet widely appreciated and
properly understood \cite{schleich}.

\nd Quoting verbatim a poetic paragraph from \cite{schleich}
 {\it  The birth of
the time-dependent Schr\"{o}dinger equation was perhaps not unlike
the birth of a river. Often, it is difficult to locate uniquely
its spring despite the fact that signs may officially mark its
beginning. Usually, many bubbling brooks and streams merge
suddenly to form a mighty river. In the case of quantum mechanics,
there are so many convincing experimental results that many of the
major textbooks do not really motivate the subject}.

\nd In this paper we uncover the rather surprising fact that the
Schr\"{o}dinger and Klein-Gordon equations can both be derived
from an hypergeometric differential equation. The same applies to
non linear generalizations of these equations, such as the ones
recently proposed, in ad hoc fashion, by  Nobre, Rego-Monteiro,
and Tsallis (NRT) \cite{tp3} (see Appendix B).

\nd The paper is organized as follows. Section 2.1 deal with the
hypergeometric  Schr\"{o}dinger derivation, Section 2.2 with the
Klein Gordon one, and Section 3 with the non linear
Schr\"{o}dinger equation, that turns out to be different from the
NRT equation. Section 4 illustrates the later equation with a wave
packet example. Section 5 hypergeometrically derives a non linear
Klein-Gordon equation, that does coincide with the NRT one. Some
conclusions are drawn in Section 6, and special details are given
in Appendixes A. B, and C.

\section{Hypergeometric derivations: Schr\"{o}dinger and Klein-Gordon equations}

\setcounter{equation}{0}

\nd We start our considerations by showing that both
Schr\"{o}dinger's and  Klein-Gordon's equations can be derived
from the differential equation satisfied by the confluent
hypergeometric  function's $\phi$ without further ado.

\subsection{Schr\"{o}dinger's wave equation}

\nd We encounter on \cite{tp5} the differential equation

\begin{equation}
\label{er1.1} U''(z)+U'(z)+\left[\frac {\lambda} {z}+\left (\frac
{1} {4} -\mu^2\right)\frac {1} {z^2}\right]U(z)=0,
\end{equation}
and a solution is

\begin{equation}
\label{er1.2} U(z)=z^{\frac {1} {2}+\mu}e^{-z}\phi(\frac {1} {2}+
\mu-\lambda,2\mu+1;z),
\end{equation}
where $\phi$ is the confluent hypergeometric function. Appealing
to the change of variables
\begin{equation}
\label{er1.3} 2\mu+1=b\;\;\;\;\;\frac {1} {2}+\mu-\lambda,
\end{equation}
Eqs.  (\ref{er1.1}) and (\ref{er1.2}) become
\begin{equation}
\label{er1.4} U''(z)+U'(z)+\left[\left(\frac {b} {2}-a\right)
\frac {1} {z}-\left (\frac {b^2-2b} {4} \right)\frac {1}
{z^2}\right]U(z)=0,
\end{equation}

\begin{equation}
\label{er1.5} U(z)=z^{\frac {b} {2}}e^{-z}\phi(a,b.z).
\end{equation}

Introducing solution (\ref{er1.5}) into (\ref{er1.4}) we ascertain
that the differential equation satisfied by  $\phi$ is

\begin{equation}
\label{er1.6} z\phi ''(a,b;z)+(b-z)\phi '(a,b;z)-a\phi(a,b;z)=0.
\end{equation}
For the special instance  $a=b$ we obtain  (see \cite{tp6})

\begin{equation}
\label{er1.7} \phi(a,a;z)=e^z,
\end{equation}
and (\ref{er1.5})) now becomes  (for $a=b$)

\begin{equation}
\label{er1.8} z\phi ''(a,a;z)+(a-z)\phi '(a,a;z)-a\phi(a,a;z)=0.
\end{equation}
Set  $z$
\begin{equation}
\label{er1.9} z=\frac {i} {\hbar} (px-Et)\;\;{\rm
and}\;\;\;E=\frac {p^2} {2m}.
\end{equation}
Then, on account of  (\ref{er1.7}) one has

\begin{equation}
\label{er1.10} e^{\frac {i} {\hbar}(px-Et)}= \phi\left[a,a;\frac
{i} {\hbar} (px-Et)\right]=\phi.
\end{equation}
The associated confluent hypergeometric differential equation is
now

\begin{equation}
\label{er1.11} \frac {i} {\hbar} (px-Et)\phi ''+\left[a-\frac {i}
{\hbar} (px-Et)\right] \phi '-a\phi=0,
\end{equation}
and applying the identities
\begin{equation}
\label{er1.12} \phi ''=-\frac {\hbar^2} {p^2} \frac
{\partial^2\phi} {\partial x^2}\;\;\;\;\; \phi '=\frac {i\hbar}
{E} \frac {\partial\phi} {\partial t}=\phi,
\end{equation}
together with a bit of algebraic manipulation, we see that
(\ref{er1.11}) adopts the appearance
\begin{equation}
\label{er1.13} -\frac {\hbar^2} {p^2} \frac {\partial^2\phi}
{\partial x^2}- \frac {i\hbar} {E} \frac {\partial\phi} {\partial
t}=0,
\end{equation}
or,  equivalently,
\begin{equation}
\label{er1.14} i\hbar \frac {\partial\phi} {\partial t}= -\frac
{\hbar^2} {2m} \frac {\partial^2\phi} {\partial x^2}.
\end{equation}
Since  $H_0=-\frac {\hbar^2} {2m} \frac {\partial^2} {\partial
x^2}$ we can finally write
\begin{equation}
\label{er1.15} i\hbar \frac {\partial\phi} {\partial t}=H_0\phi,
\end{equation}
i.e.,  Schr\"{o}dinger's free particle equation. For an arbitrary
Hamiltonian $H$, (\ref{er1.15}) may be generalized to
\begin{equation}
\label{er1.16} \frac i\hbar \frac {\partial\phi} {\partial
t}=H\phi,
\end{equation}
the usual  Schr\"{o}dinger equation (SE). Thus, we obtain
Schr\"{o}dinger's wave equation directly from the hypergeometric
differential equation and then, by suitable generalization, the
usual SE.

\subsection{Klein-Gordon's equation}

\nd We start with

\begin{equation}
\label{er1.17} z=i(kx-\omega t)\;\;\;\;\; \omega^2=k^2c^2+\frac
{m^2c^4} {\hbar^2},
\end{equation}
and write

\begin{equation}
\label{er1.18} e^{i(kx-\omega t)}= \phi\left[a,a;i(kx-\omega
t)\right]=\phi.
\end{equation}
The operating confluent hypergeometric differential equation is
here

\begin{equation}
\label{er1.19} i(kx-\omega t)\phi ''+\left[a-i(kx-\omega t)\right]
\phi '-a\phi=0,
\end{equation}
so that, by appeal to the identities

\begin{equation}
\label{er1.20} \phi ''=-\frac {1} {k^2} \frac {\partial^2\phi}
{\partial x^2}\;\;\;\;\; \phi '=\frac {i} {\omega} \frac
{\partial\phi} {\partial t}=\phi,
\end{equation}
we get for (\ref{er1.19}):

\begin{equation}
\label{er1.21} -\frac {1} {k^2} \frac {\partial^2\phi} {\partial
x^2}- \frac {i} {\omega } \frac {\partial\phi} {\partial t}=0.
\end{equation}
Since
\begin{equation}
\label{er1.22} \frac {\partial \phi} {\partial t}= \frac {i}
{\omega}\frac {\partial^2\phi} {\partial t^2},
\end{equation}
 (\ref{er1.21}) becomes
\begin{equation}
\label{er1.23} -\frac {1} {k^2} \frac {\partial^2\phi} {\partial
x^2}+ \frac {i} {\omega^2} \frac {\partial^2\phi} {\partial
t^2}=0.
\end{equation}
Using now the equality

\begin{equation}
\label{er1.24} \frac {\partial^2\phi} {\partial x^2}= -k^2\phi,
\end{equation}
and a little algebra we arrive at

\begin{equation}
\label{er1.25} \frac {1} {c^2}\frac {\partial^2\phi} {\partial
t^2}- \frac {\partial^2\phi} {\partial x^2}+ \frac {m^2c^2}
{\hbar^2}\phi=0,
\end{equation}
the desired  Klein-Gordon equation.

\section{A non linear Schr\"{o}dinger equation}

\setcounter{equation}{0}

\nd  The above procedure can be generalized by appealing to the
more general hypergeometric function $F$. This $F-$treatment has
been worked out by us in \cite{PR}, {\it only for the
Schr\"{o}dinger case and without reference} to the present
confluent instance, originally developed in Section 2 above. We
reconsider this non linear generalization here for completeness's
sake, in order to better appreciate the workings of our
hypergeometric approach. See more details in Appendix A.

\nd As shown in \cite{PR}, one can write
\begin{equation}
\label{eq2.1a} F(-\alpha,\gamma;\gamma;-z)=(1+z)^{\alpha}.
\end{equation}
It follows that for Tsallis' imaginary q-exponential function (see
its definition in \cite{PR}-\cite{tp3}) one has
\begin{equation}
\label{eq2.2a} \left[1+\frac {i} {\hbar}(1-q)(px-Et)\right]^{\frac
{1} {1-q}}= F\left[\frac {1} {q-1},\gamma;\gamma;\frac {i} {\hbar}
(q-1)(px-Et)\right]\equiv F,
\end{equation}
where $E=\frac {p^2} {2m}$. According to  \cite{tp2}, $F$
satisfies
\begin{equation}
\label{eq2.3a} z(1-z)F''(\alpha,\beta;\gamma;z)+
[\gamma-(\alpha+\beta+1)z]F'(\alpha,\beta;\gamma;z)- \alpha\beta
F(\alpha,\beta;\gamma;z)=0.
\end{equation}
After following developments detailed in Appendix A one arrives at
\begin{equation}
\label{eq2.17} i\hbar \frac {\partial}{\partial t} \left[\frac
{F(x,t)} {F(0,0)}\right]^q= H_0 \left[\frac {F(x,t)}
{F(0,0)}\right].
\end{equation}
Generalizing to arbitrary $H$ we have
\begin{equation}
\label{eq2.18} i\hbar \frac {\partial}{\partial t} \left[\frac
{\psi(x,t)} {\psi(0,0)}\right]^q= H \left[\frac {\psi(x,t)}
{\psi(0,0)}\right].
\end{equation}
Differences and similarities between these equations and those
obtained by Nobre, Rego-Monteiro, and Tsallis (NRT) \cite{tp3} are
fully discussed in \cite{PR}. Some NRT details are reviewed in
Appendix B.

\section{The q-Gaussian wave packet}

\setcounter{equation}{0}

\nd We pass now to discuss an important solution of
(\ref{eq2.18}): the wave packet. Details of the ordinary case are
summarized in Appendix C.  The pertinent analysis for the NRT
equation has been given in \cite{curilef}. Setting $\psi(0,0)=1$
one has
\begin{equation}
\label{er3.1}
i\hbar\frac {\partial\psi^q}
{\partial t}=
-\frac {\hbar^2} {2m}
\frac {\partial^2\psi}
{\partial x^2}
\end{equation}
Following \cite{curilef}, we propose as a solution
\begin{equation}
\label{er3.2}
\psi=\left\{1+(q-1)\left[a(t)x^2+b(t)x+c(t)\right]\right\}^{ \frac
{1} {1-q}},
\end{equation}
where  $a, b,$ and $c$ are temporal functions to be determined.
From  $\psi(0,0)=1$ one has $c(0)=0$. Deriving $\psi^q$ with
respect to time we find
\begin{equation}
\label{er3.3} \frac {\partial\psi^q} {\partial t}=-q
\left[\dot{a}(t)x^2+\dot{b}(t)x+\dot{c}(t)\right] \psi^{q-1}.
\end{equation}
We then look for the second $\psi-$ derivative with respect to
$x$:
\begin{equation}
\label{er3.4} \frac {\partial^2\psi} {\partial x^2}=
\left[2(q+1)a^2x^2+2(q+1)abx+b^2 -2(q-1)ac-2a\right] \psi^{q-1}.
\end{equation}
Introducing these two results into  (\ref{er3.1}) we are led to a
non linear system for   $a, b,$ and $c$
\begin{equation}
\label{er3.5} imq\dot{a}=\hbar(q+1)a^2,
\end{equation}
\begin{equation}
\label{er3.6} imq\dot{b}=\hbar(q+1)ab,
\end{equation}
\begin{equation}
\label{er3.7} 2imq\dot{c}=\hbar\left[qb^2-2(q-1)ac-2a\right].
\end{equation}
This system's solution is given by
\begin{equation}
\label{er3.8} a(t)=\frac {mq} {i\hbar(q+1)t+mq\alpha},
\end{equation}
\begin{equation}
\label{er3.9} b(t)=\frac {1} {\beta} \frac {1}
{[i\hbar(q+1)t+mq\alpha]},
\end{equation}
\[c(t)=(mq\alpha)^{\frac {1-q} {1+q}}
\left(\frac {1} {q-1}-\frac {1} {4m^2q^2\beta^2\alpha}\right)
\left[i\hbar(q+1)t+mq\alpha\right]^{\frac {q-1} {q+1}}+\]
\begin{equation}
\label{er3.10} \frac {1} {4mq\beta^2[i\hbar(q+1)t+mq\alpha]}+
\frac {1} {1-q},
\end{equation}
where $\alpha$ and $\beta$ are constants to be fixed according to
initial or boundary conditions for (\ref{er3.1}).

\nd It is straightforward to prove that equations (\ref{er3.8}),
(\ref{er3.9}) and (\ref{er3.10}) are transformed into equations
(\ref{eq6.8}), (\ref{eq6.9}) and (\ref{eq6.10}) of Appendix C for
$q\rightarrow 1$. Thus the q-Gaussian wave packet transforms into
the Gaussian wave packet when $q\rightarrow 1$. Since our Eq.
(\ref{eq2.17}) is different from NRT's non linear one, so are also
their associated wave packet solutions.

\section{A hypergeometric-generated non-linear Klein-Gordon Equation }

\setcounter{equation}{0}

\nd We derive now in hypergeometric fashion a non linear KG
equation satisfied by the q-exponential function. Let
$z=i(q-1)(kx-\omega t)$. Then, \ben \label{er2.1} & 1-z =
\left[1+i(1-q)(kx-\omega t)\right]^{\frac {1} {1-q}}=\cr & =
F\left[\frac {1} {q-1},\gamma;\gamma;i (q-1)(kx-\omega
t)\right]\equiv F. \een Recourse to the equalities
\begin{equation}
\label{er2.2} F''=-\frac {1} {k^2(q-1)^2} \frac{\partial^2F}
{\partial x^2},
\end{equation}
\begin{equation}
\label{er2.3} F'=-\frac {1} {i\omega(q-1)} \frac {\partial F}
{\partial t}= -\frac {F^{(1-q)}} {\omega^2q(q-1)} \frac
{\partial^2F} {\partial t^2},
\end{equation}
\begin{equation}
\label{er2.4} F=\frac {i} {\omega}F^{(1-q)} \frac {\partial F}
{\partial t},
\end{equation}
allow one to obtain, via (\ref{eq2.3}):
\[\frac {z(1-z)} {k^2(q-1)^2}\frac {\partial^2F} {\partial x^2}+
\left[\gamma-\left(\frac {q} {q-1}+\gamma\right)z\right]
\frac {1} {i\omega(q-1)}\frac {\partial F} {\partial t}+\]
\begin{equation}
\label{er2.5} \frac {i\gamma} {\omega(q-1)}F^{(1-q)} \frac
{\partial F} {\partial t}=0,
\end{equation}
or, equivalently,
\[\frac {z(1-z)} {k^2(q-1)^2}\frac {\partial^2F} {\partial x^2}+
\left[\gamma(1-z)-\left(\frac {qz} {q-1}\right)\right]
\frac {1} {i\omega(q-1)}\frac {\partial F} {\partial t}+\]
\begin{equation}
\label{er2.6} \frac {i\gamma} {\omega(q-1)}F^{(1-q)} \frac
{\partial F} {\partial t}=0.
\end{equation}
Now, from (\ref{er2.1}), one has  $1-z=F^{(1-q)}$ so that
(\ref{er2.6}) adopts the appearance
\[\frac {zF^{(1-q)}} {k^2(q-1)}\frac {\partial^2F} {\partial x^2}+
\left[\gamma F^{(1-q)}-\left(\frac {qz} {q-1}\right)\right]
\frac {1} {i\omega}\frac {\partial F} {\partial t}+\]
\begin{equation}
\label{er2.7} \frac {i\gamma} {\omega}F^{(1-q)} \frac {\partial F}
{\partial t}=0,
\end{equation}
and, simplifying terms,
\begin{equation}
\label{er2.8} \frac {zF^{(1-q)}} {k^2(q-1)} \frac {\partial^2F}
{\partial x^2}+ \frac {iqz} {\omega(q-1)} \frac {\partial F}
{\partial t}=0,
\end{equation}
entailing
\begin{equation}
\label{er2.9} \frac {F^{(1-q)}} {k^2} \frac {\partial^2F}
{\partial x^2}+ \frac {iq} {\omega} \frac {\partial F} {\partial
t}=0.
\end{equation}
Using here  (\ref{er2.3}) re reach
\begin{equation}
\label{er2.10} \frac {F^{(1-q)}} {k^2} \frac {\partial^2F}
{\partial x^2}- \frac {F^{(1-q)}} {\omega^2} \frac {\partial^2 F}
{\partial t^2}=0,
\end{equation}
which, simplifying the common factor $F^{1-q}$ in (\ref{er2.10}),
yields
\begin{equation}
\label{er2.11} \frac {1} {k^2} \frac {\partial^2F} {\partial x^2}-
\frac {1} {\omega^2} \frac {\partial F} {\partial t}=0.
\end{equation}
Since, additionally, one has
\begin{equation}
\label{er2.12} \frac {\partial^2F} {\partial x^2}=- k^2qF^{2q-1)},
\end{equation}
some algebra leads to
\begin{equation}
\label{er2.13} \frac {1} {c^2}\frac {\partial^2F} {\partial t^2}-
\frac {\partial^2F} {\partial x^2}+ \frac {qm^2c^2}
{\hbar^2}F^{(2q-1)}=0.
\end{equation}
If  $\phi$ is given by
\begin{equation}
\label{er2.14} \phi(x,t)=A\left[1+i(1-q)(kx-\omega
t)\right]^{\frac {1} {1-q}},
\end{equation}
we find, via  (\ref{er2.13}),

\begin{equation} \label{er2.15} \frac {1} {c^2}\frac {\partial^2}
{\partial t^2} \left[\frac {\phi(x,t)} {\phi(0,0)}\right]- \frac
{\partial^2} {\partial x^2} \left[\frac {\phi(x,t)}
{\phi(0,0)}\right]+ \frac {qm^2c^2} {\hbar^2} \left[\frac
{\phi(x,t)} {\phi(0,0)}\right]^{(2q-1)}=0,
\end{equation}
that in  $n$ dimensions becomes

\begin{equation} \label{er2.16}
\frac {1} {c^2}\frac {\partial^2} {\partial t^2} \left[\frac
{\phi(\vec{x},t)} {\phi(0,0)}\right]- \nabla^2 \left[\frac
{\phi(\vec{x},t)} {\phi(0,0)}\right]+ \frac {qm^2c^2} {\hbar^2}
\left[\frac {\phi(\vec{x},t)} {\phi(0,0)}\right]^{(2q-1)}=0,
\end{equation}
where
\begin{equation}
\label{er2.17} \phi(\vec{x},t)=A\left[1+i(1-q)
(\vec{k}\cdot\vec{x}-\omega t)\right]^{\frac {1} {1-q}}.
\end{equation}
Eq. (\ref{er2.16}) and its two dimensional case  (\ref{er2.15})
coincides with the non lineal Klein-Gordon equation advanced by
NRT in \cite{tp3}.

\section{Conclusions}

\setcounter{equation}{0} \nd In this work we have uncovered a
rather surprising fact. Both  that the Schr\"{o}dinger and
Klein-Gordon equations can be derived from an hypergeometric
differential equation. The same procedure can be applied  to non
linear generalizations of these equations, such as the ones
recently proposed by  Nobre, Rego-Monteiro, and Tsallis (NRT)
\cite{tp3}.

\section{Acknowledgements}

\setcounter{equation}{0}

We acknowledge very fruitful discussion with Prof. A. R. Plastino.

\newpage

\renewcommand{\thesection}{\Alph{section}}

\renewcommand{\theequation}{\Alph{section}.\arabic{equation}}

\setcounter{section}{1}

\section*{Appendix A}

\subsection*{Non linear Schr\"{o}dinger Eq. of \cite{PR} (review)}

\setcounter{equation}{0}

\nd  The procedure of Section 2 can be generalized by appealing to
the more general hypergeometric function $F$. This $F-$treatment
has been previously developed by us in \cite{PR}, {\it only for
the Schr\"{o}dinger case and without reference} to the  confluent
instance. We review this generalization here for completeness's
sake. \vskip 4mm

\nd As shown in \cite{PR}, one can write
\begin{equation}
\label{eq2.1} F(-\alpha,\gamma;\gamma;-z)=(1+z)^{\alpha}.
\end{equation}
It follows that for Tsallis' imaginary q-exponential function (see
its definition in \cite{PR}-\cite{tp3}) one has
\begin{equation}
\label{eq2.2} \left[1+\frac {i} {\hbar}(1-q)(px-Et)\right]^{\frac
{1} {1-q}}= F\left[\frac {1} {q-1},\gamma;\gamma;\frac {i} {\hbar}
(q-1)(px-Et)\right]\equiv F,
\end{equation}
where $E=\frac {p^2} {2m}$. According to  \cite{tp2}, $F$
satisfies
\begin{equation}
\label{eq2.3} z(1-z)F''(\alpha,\beta;\gamma;z)+
[\gamma-(\alpha+\beta+1)z]F'(\alpha,\beta;\gamma;z)- \alpha\beta
F(\alpha,\beta;\gamma;z)=0.
\end{equation}
For the particular case (\ref{eq2.2}), this last expression adopts
the appearance
\[\frac {i} {\hbar}(q-1)(px-Et)\left[
1-\frac {i} {\hbar}(q-1)(px-Et)\right] F''+\]
\begin{equation}
\label{eq2.4} \left[\gamma-\left(\frac {1} {q-1}+\gamma+1\right)
\frac {i} {\hbar}(q-1)(px-Et)\right] F'-\frac {\gamma} {q-1}F=0.
\end{equation}
 Accordingly, one can deduce a relationship between $\dot F$
 (time) and $F'$ (space) [see \cite{PR} for details]

\[F'\left[\frac {1} {q-1},\gamma;\gamma;\frac {i} {\hbar}
(q-1)(px-Et)\right]=\]
\begin{equation}
\label{eq2.5} \frac {i\hbar} {(q-1)E}\frac {\partial} {\partial t}
F\left[\frac {1} {q-1},\gamma;\gamma;\frac {i} {\hbar}
(q-1)(px-Et)\right].
\end{equation}
In similar fashion one obtains \cite{tp2}

\[F''\left[\frac {1} {q-1},\gamma;\gamma;\frac {i} {\hbar}
(q-1)(px-Et)\right]=\]
\begin{equation}
\label{eq2.6} -\frac {\hbar^2} {(q-1)^2p^2}\frac {\partial^2}
{\partial x^2} F\left[\frac {1} {q-1},\gamma;\gamma;\frac {i}
{\hbar} (q-1)(px-Et)\right].
\end{equation}
Replacing  (\ref{eq2.5}) and (\ref{eq2.6}) into (\ref{eq2.4}), the
later becomes
\[-\frac {i} {\hbar}(q-1)(px-Et)\left[
1-\frac {i} {\hbar}(q-1)(px-Et)\right] \frac {\hbar^2}
{(q-1)^2p^2}\frac {\partial^2F} {\partial x^2} +\]
\[\left[\gamma-\left(\frac {1} {q-1}+\gamma+1\right)
\frac {i} {\hbar}(q-1)(px-Et)\right] \frac {i\hbar} {(q-1)E}\frac
{\partial F} {\partial t}\]
\begin{equation}
\label{eq2.7} -\frac {\gamma} {q-1} F=0,
\end{equation}
that, in turn, can be rewritten as

\[-\frac {i} {\hbar}(q-1)(px-Et)\left[
1-\frac {i} {\hbar}(q-1)(px-Et)\right]\times \frac {\hbar^2}
{(q-1)m^2}\frac {\partial^2F} {\partial x^2}+\]\begin{equation}
\label{eq2.8} \left[\gamma-\left(\frac {1} {q-1}+\gamma+1\right)
\frac {i} {\hbar}(q-1)(px-Et)\right] i\hbar\frac {\partial F}
{\partial t}-\gamma EF=0.
\end{equation}
Also, we obtain from (\ref{eq2.2})
\[-\gamma E
F\left[\frac {1} {q-1},\gamma;\gamma;\frac {i} {\hbar}
(q-1)(px-Et)\right]=\]
\[-i\hbar\gamma
\left\{F\left[\frac {1} {q-1},\gamma;\gamma;\frac {i} {\hbar}
(q-1)(px-Et)\right]\right\}^{(1-q)}\times\]
\begin{equation}
\label{eq2.9} \frac {\partial} {\partial t} F\left[\frac {1}
{q-1},\gamma;\gamma;\frac {i} {\hbar} (q-1)(px-Et)\right].
\end{equation}
Using n (\ref{eq2.9}), (\ref{eq2.8}) adopts the appearance
\[-\frac {\hbar^2} {2m(q-1)}
\left[1-F^{(1-q)}\right]F^{(1-q)} \frac {\partial^2} {\partial
x^2}F+\]
\[i\hbar\left\{\gamma+\left(\frac {1} {q-1}+
\gamma+1\right)\left[F^{(1-q)}-1\right]\right\} \frac {\partial}
{\partial t} F-\]
\begin{equation}
\label{eq2.11} i\hbar\gamma F^{(1-q)}\frac {\partial} {\partial t}
F=0,
\end{equation}
and, after simplifying,
\begin{equation}
\label{eq2.12} -\frac {\hbar^2} {2m} F^{(1-q)} \frac {\partial^2}
{\partial x^2}F-i\hbar q\frac {\partial} {\partial t}F=0,
\end{equation}
that can be recast as
\begin{equation}
\label{eq2.13} i\hbar q\frac {\partial}{\partial t}F= F^{(1-q)}H_0
F,
\end{equation}
where$H_0$ is the free particle Hamiltonian. Note that for  $q=1$
 things`properly reduce to  Schr\"{o}dinger's wave equation.
If, instead of  (\ref{eq2.2}) we have
\begin{equation}
\label{eq2.14} F(x,t)=A \left[1+\frac {1}
{\hbar}(1-q)(px-Et)\right]^{\frac {1} {1-q}},
\end{equation}
then  $F(0,0)=A$ and  (\ref{eq2.13}) becomes
\begin{equation}
\label{eq2.15} i\hbar q\frac {\partial}{\partial t} \left[\frac
{F(x,t)} {F(0,0)}\right]= {\left[\frac {F(x,t)}
{F(0,0)}\right]}^{(1-q)}H_0 \left[\frac {F(x,t)} {F(0,0)}\right],
\end{equation}
or, equivalently,
\begin{equation}
\label{eq2.16} i\hbar q {\left[\frac {F(x,t)}
{F(0,0)}\right]}^{(q-1)} \frac {\partial}{\partial t} \left[\frac
{F(x,t)} {F(0,0)}\right]= H_0 \left[\frac {F(x,t)}
{F(0,0)}\right],
\end{equation}
that can in turn be written as
\begin{equation}
\label{eq2.17a} i\hbar \frac {\partial}{\partial t} \left[\frac
{F(x,t)} {F(0,0)}\right]^q= H_0 \left[\frac {F(x,t)}
{F(0,0)}\right].
\end{equation}
Generalizing to arbitrary $H$ we have
\begin{equation}
\label{eq2.18a} i\hbar \frac {\partial}{\partial t} \left[\frac
{\psi(x,t)} {\psi(0,0)}\right]^q= H \left[\frac {\psi(x,t)}
{\psi(0,0)}\right].
\end{equation}
Differences and similarities between these equations and those
obtained by Nobre, Rego-Monteiro, and Tsallis (NRT) \cite{tp3} are
fully discussed in \cite{PR}.

\setcounter{section}{2}

\section*{Appendix B}

\subsection*{The NRT equation}

\setcounter{equation}{0}

We review here the NRT equation for the free particle in order to
make this paper self-contained. It reads \cite{tp3}

\begin{equation}
\label{eq4.1} i\hbar (2-q) \frac {\partial}{\partial t}
\left[\frac {\psi(\vec{x},t)} {\psi(0,0)}\right]= H_0 \left[\frac
{\psi(\vec{x},t)} {\psi(0,0)}\right]^{2-q}.
\end{equation}
Setting  $\phi=\psi^{2-q}$ we are led to
\begin{equation}
\label{eq4.2} i\hbar (2-q) \frac {\partial}{\partial t}
\left[\frac {\phi(\vec{x},t)} {\phi(0,0)}\right]^{ \frac {1}
{1-q}}= H_0 \left[\frac {\phi(\vec{x},t)} {\phi(0,0)}\right],
\end{equation}
and generalizing to arbitrary $H$ one writes
\begin{equation}
\label{eq4.3} i\hbar (2-q) \frac {\partial}{\partial t}
\left[\frac {\phi(\vec{x},t)} {\phi(0,0)}\right]^{ \frac {1}
{1-q}}= H \left[\frac {\phi(\vec{x},t)} {\phi(0,0)}\right],
\end{equation}
that in $\psi$ terms becomes
\begin{equation}
\label{eq4.4} i\hbar (2-q) \frac {\partial}{\partial t}
\left[\frac {\psi(\vec{x},t)} {\psi(0,0)}\right]= H \left[\frac
{\psi(\vec{x}x,t)} {\psi(0,0)}\right]^{2-q}.
\end{equation}

\setcounter{section}{3}

\section*{Appendix C}

\subsection*{The ordinary Gaussian wave packet}

\setcounter{equation}{0}

\nd We want to tackle
\begin{equation}
\label{eq6.1} i\hbar\frac {\partial\psi(x,t)} {\partial t}= -\frac
{\hbar^2} {2m} \frac {\partial^2\psi(x,t)} {\partial x^2},
\end{equation}
via the Gaussian packet
\begin{equation}
\label{eq6.2} \psi(x,t)=e^{-[a(t)x^2+b(t)x+c(t)]},
\end{equation}
with the initial  condition $\psi(0,0)=1$, entailing $c(0)=0$.
After a time derivative we get
\begin{equation}
\label{eq6.3} \frac {\partial\psi(x,t)} {\partial t}=
-[\dot{a}(t)x^2+\dot{b}(t)x+\dot{c}(t)]\psi(x,t).
\end{equation}
The spatial second derivative yields
\begin{equation}
\label{eq6.4} \frac {\partial^2\psi(x,t)} {\partial x^2}=
-[4a^2(t)x^2+4a(t)b(t)x+ b^2(t)-2a(t)]\psi(x,t).
\end{equation}
Replacing (\ref{eq6.3}) and (\ref{eq6.4}) into (\ref{eq6.1}) we
obtain
\begin{equation}
\label{eq6.5} \dot{a}=\frac {2\hbar} {im} a^2,
\end{equation}
\begin{equation}
\label{eq6.6} \dot{b}=\frac {2\hbar} {im} ab,
\end{equation}
\begin{equation}
\label{eq6.7} \dot{c}=\frac {\hbar} {2im}(b^2-2a),
\end{equation}
whose solution reads
\begin{equation}
\label{eq6.8} a(t)=\frac {m} {2i\hbar t+m\alpha},
\end{equation}
\begin{equation}
\label{eq6.9} b(t)=\frac {1} {\beta}\frac {m} {[2i\hbar
t+m\alpha]},
\end{equation}
\[c(t)=\frac {1} {4m\beta^2[2i\hbar t+m\alpha]}+
\frac {1} {2} \ln(2i\hbar t+m\alpha)-
\frac {1} {4m^2\beta^2\alpha}-\]
\begin{equation}
\label{eq6.10} \frac {1} {2}\ln(m\alpha),
\end{equation}
where for  $c$ one has $c(0)=0$.

\end{document}